\input amstex

\documentstyle{amsppt} \magnification1200 \TagsOnRight
\NoBlackBoxes
\pagewidth{30pc} \pageheight{47pc} 

\def\today{September 4, 1998; revised January 26, 1999}

\def\SU{\operatorname {SU}}

\def\SO{\operatorname {SO}}
\def\OO{\operatorname {O}}

\def\Tr{\operatorname {Tr}}

\def\R{{\Bbb R}}

\def\d{{\text{d}}}

\topmatter
\title The asymptotics of an amplitude for the 4-simplex\endtitle
\author John W. Barrett\\
 Ruth M Williams\endauthor
\date\today\enddate
\address
Center for Gravitational Physics and Geometry,
Department of Physics,
The Pennsylvania State University,
104 Davey Laboratory,
University Park, PA 16802--6300,
USA
\endaddress
\email jwb\@maths.nott.ac.uk \endemail
\address
 DAMTP,
 Silver Street,
 Cambridge CB3 9EW, UK
\endaddress
\email R.M.Williams\@damtp.cam.ac.uk \endemail

\abstract
An expression for the oscillatory part of an asymptotic formula for the  relativistic spin network amplitude for a 4-simplex is given.
The amplitude depends on specified areas for each two-dimensional face in 
the 4-simplex. The asymptotic formula has 
a contribution from each flat Euclidean metric on the 4-simplex which 
agrees with the given areas. The oscillatory part of each contribution is 
determined by the Regge calculus Einstein action for that geometry.
\endabstract

 \endtopmatter

\document
\head Introduction \endhead

The purpose of this paper is to give a physical interpretation of the
function of 10 balanced representations of the Lie group $\SO(4)$
introduced in \cite{Barrett and Crane 1998}. This value of the function is a 
real number calculated from a relativistic spin network associated to a
4-simplex. We call it a symbol, by analogy with the terminology of
6j-symbols for $\SU(2)$\footnote{It is tempting to call this a 10j-sym\-bol, 
but the terminology Nj-sym\-bol already has a specific meaning, and 
the symbol is not one of these.}. 

Each balanced representation is determined by a non-negative 
half-integer $j$, the spin.
The 10 balanced representations are associated to the 10 triangles of the
4-simplex.

In this paper we determine the leading part of an asymptotic formula for
the symbol, inspired by the corresponding formula for a $6j$-symbol given
by \cite{Ponzano and Regge 1968}. The formula has a contribution from each
metric on the 4-simplex for which the area of the triangle is given by
$2j+1$, where $j$ is the spin label for that triangle. The phase factor
for each contribution is determined by the Regge calculus formula for the
Einstein action of the 4-simplex. An argument connecting the closely related balanced 
15j-symbol and the Einstein action was given by \cite{Crane and Yetter 1997}.

The general context for the symbol as an amplitude in a state sum model was introduced 
in \cite{Barrett and Crane 1998}, and developed in \cite{Baez 1998}. Some more 
background is explained in the Penn State lecture \cite{Barrett 1998b}.

\head The symbol \endhead

The symbol was originally defined in terms of a relativistic spin network 
evaluation in \cite{Barrett and Crane 1998}. For the classical Lie group 
considered here, this is defined by taking 
the value of the 15j-symbol for SU(2) which is associated to the 
4-simplex with 5 additional spins specified at tetrahedra. The square of 
this number is then summed over the 5 additional spins, with appropriate 
weights. 

The relativistic spin
network evaluation was shown to be given by an integral over copies of the
Lie group $\SU(2)$ in \cite{Barrett 1998a}. (In that paper, the 
integer $n=2j$ was called the spin.)  This definition is used as the starting
point for this paper. 

The five tetrahedra in the 4-simplex are numbered by $k=1,2,\ldots 5$, 
and the triangles are indexed by the pair $k,l$ of tetrahedra which 
intersect on the triangle. The 10 spins are thus $\{j_{kl}| k<l \}$.

The matrix representing an element 
$g\in\SU(2)$ in the irreducible representation of spin $j_{kl}$ belonging 
to a triangle is denoted $\rho_{kl}(g)$.

A variable $h_k\in\SU(2)$ is assigned to each tetrahedron $k$. The invariant
$I\in \R$ is defined by integrating a function of these variables over 
each copy of $\SU(2)$. 

The evaluation of the relativistic spin network is
$$ I=(-1)^{\sum_{k<l} 2j_{kl}} \int_{h\in SU(2)^5} 
\prod_{k<l} \Tr \rho_{kl}\left(h_{k}^{\strut} h_{l}^{-1}\right).$$
The integration measure is the Haar measure on each of the five 
copies of $\SU(2)$, normalised to total volume 1.  

The geometrical interpretation of this formula given in \cite{Barrett 
1998a} is that since the manifold $\SU(2)$ is isomorphic to $S^3$ by 
$$x\mapsto\pmatrix x_0+ix_1 & x_2+i x_3\\-x_2+ix_3& x_0-ix_1 \endpmatrix.$$
each variable $h\in\SU(2)$ can be regarded as a
unit vector in $\R^4$. Further, this unit vector can be regarded as the
normal vector to a 3-dimensional hyperplane in $\R^4$ through the 
origin. The idea is to interpret this as the hyperplane in 
which the tetrahedron lies in a geometrical simplex in $\R^4$. Similar variables were introduced in a first order version of classical Regge calculus by \cite{Caselle, D'Adda and Magnea 1989}.

The weight for one triangle $kl$
is a function of the angle $\phi$ between the two unit vectors $h_{k}$
and $h_{l}$ in $\R^4$,
$$\Tr \rho\left(h_{k}^{\strut}
h_{l}^{-1}\right)={\sin(2j+1)\phi\over\sin\phi}. $$
This angle, defined by $\cos\phi=h_k\cdot h_l$, is the exterior angle between the two hyperplanes. Since the 
sum of the spins at a tetrahedron is an integer, the integrand is 
unchanged if 
$h$ is replaced by $-h$ at one tetrahedron. Thus the integrand does not 
register the orientation of the hyperplanes.

The set of five hyperplanes determines a geometric 4-simplex in $\R^4$ up
to overall scale and parallel translation, assuming the generic
case where any four of the unit vectors are linearly independent. Moving
one of the hyperplanes parallel to itself away from the origin, defines a
geometric 4-simplex as the space bounded by the hyperplanes. The distance
the hyperplane is moved from the origin is undetermined, so the overall
scale of the simplex is undetermined. This possible scaling includes negative
scaling factors, which invert the 4-simplex. 
 
In this way, the integration can be regarded as an integration over the 
set of geometric 4-simplexes modulo isometries and scaling. The 
isometries are products of translations 
and transformations by elements of $\OO(4)$. This set of 4-simplexes is 
parameterised by 
the ten angles $\phi$, which are subject to one constraint equation. This 
is the same as the space of edge lengths for the simplex, modulo scaling 
all 10 lengths simultaneously.

Note that the scaling by a factor $-1$ does not change the orientation, 
so the geometric simplexes up to scaling are oriented simplexes. However 
the value of the integrand does not depend on this orientation.

\head Asymptotics \endhead

For large values of the spins, the asymptotic value of the integral can 
be calculated using the method of stationary phase. The important terms 
are the  $\sin(2j+1)\phi$, as these depend on the asymptotic 
parameters, the spins $j$. These are expanded as exponentials to 
apply the stationary phase method, 
$$\sin(2j+1)\phi=
{1\over 2i}\bigl(\exp{ i(2j+1)\phi} - \exp{ -i(2j+1)\phi}\bigr).$$ 
The $\sin\phi$ factors in the denominator do 
not vary as $j\to\infty$, so can be treated as part of the 
integration measure. 

For each triangle, introduce a variable $\epsilon_{kl}$ which 
can take the values $\pm1$. Then the integral expression for $I$ can be 
written
$$ I={(-1)^{\sum_{k<l} 2j_{kl}}\over(2i)^{10}} 
\sum_{{\scriptstyle
\epsilon_{12}=\pm1\atop
\scriptstyle\ldots }\atop
\scriptstyle\epsilon_{45}=\pm1}
\int_{h\in SU(2)^5} 
 \left(\prod_{k<l}{\epsilon_{kl}\over\sin\phi_{kl}}\right) 
\exp\left({ i\sum_{k<l}\epsilon_{kl}(2j_{kl}+1)\phi_{kl}}\right) .$$ 
Now each term in this sum depends on slightly more than the angles of a 
simplex. This is because the term is not invariant under the replacement 
$h\to -h$. The geometric simplex determined by the hyperplanes has 
outward normal vectors $n_k$ for the tetrahedra, and $n_k=\pm h_k$ for 
each $k$. This means that either $\phi$ or $\pi-\phi$ are the exterior 
angles of the geometric 4-simplex.

Each integral can be split into a number of domains where these 
five signs take particular values. Now changing $h\to -h$ shows that each 
domain is equivalent to a domain where $n_k=h_k$ for all $k$, but for 
different values of the $\epsilon_{kl}$. Therefore we can assume that 
$n_k=h_k$, and so the $\phi$ are exactly the exterior angles of the 
geometric simplex. Now it is possible to compute the stationary points. 
These are the stationary points of the action
$$ S=\sum_{k<l}\epsilon_{kl}(2j_{kl}+1)\phi_{kl} .$$ 
The one constraint between the $\phi$ can be taken account of using 
Schl\"afli's differential identity [Ponzano and Regge 1968] 
$$\sum_{k<l} A_{kl}\,\d \phi_{kl}=0,$$
where the $A_{kl}$ are the areas of the triangles of one of the geometric 
4-simplexes determined by the $\phi$ (i.e., fixing a particular scale for 
the 4-simplex). This constrained variation is exactly as 
for the corresponding first order formalism for Regge calculus 
\cite{Barrett 1994}.
Using a Lagrange multiplier 
$\mu$ for this constraint,
$$\d S =\sum_{k<l}\epsilon_{kl}(2j_{kl}+1)\,\d \phi_{kl}  
= \mu \sum A_{kl}\,\d \phi_{kl},$$
and so for each triangle
$$ \epsilon_{kl}(2j_{kl}+1)= \mu A_{kl}.$$ 
Since the $\{A_{kl}\}$ are only determined up to overall scale, it is most 
convenient to take $\mu=\pm1$ to fix this overall scale. Then the result 
is that for a stationary phase point in the integrand
\roster
\item The $\epsilon_{kl}$ are either all positive or all negative.
\item The angles $\phi$ are those of a geometric 4-simplex with areas 
$A_{kl}=2j_{kl}+1$.
\item The integrand is $\exp
 i \mu S_E$, with $S_E=\sum_{k<l}A_{kl}\phi_{kl}$ 
which is the Regge calculus version of the Einstein action for a 
4-simplex, and $\mu=\pm1$.
\endroster

The last point about the Einstein action requires a little more 
explanation. The Einstein action for a 4-manifold is the integral of the 
scalar curvature on the manifold plus the integral of the mean curvature 
of the boundary. Here, the scalar curvature is zero and so the action for 
the flat 4-simplex is entirely the boundary term \cite{Hartle and Sorkin 
1981}. 

Each stationary phase point occurs $2^5$ times, so the asymptotic formula is
$$ I \sim   -{(-1)^{\sum 2j}\over 2^4} 
\left(\sum_\sigma P(\sigma) \cos \left( S_E(\sigma)+\kappa{\pi\over4}\right)\right) + D $$
The sum is over the set of metric 4-simplexes $\sigma\subset\R^4$ 
modulo isometries, which are such that the area of the triangle $kl$ is 
given by $2j_{kl}+1$. As the squares of the areas are quadratic 
polynomials in the squares of the edge lengths, and the mapping is 
non-degenerate, at least in the regular cases \cite{Barrett 1994}, this 
appears to be a finite set.

The prefactor $P$ is calculated in the usual way for stationary phase in 
terms of a determinant. This does not oscillate with the 
asymptotic parameters $j$ and so is quite different to the cosine term. 
Finally, there is also a contribution $D$ to the asymptotics of the integral 
from degenerate simplexes, when two of the $h$'s coincide. We have not 
analysed these terms.

\enddocument